\documentclass[lettersize,journal]{IEEEtran}
\usepackage[numbers]{natbib}
\usepackage{graphicx} 
\usepackage{authblk}
\usepackage[utf8]{inputenc} 
\usepackage[T1]{fontenc}    
\usepackage{hyperref}       
\usepackage{url}            
\usepackage{booktabs}       
\usepackage{amsfonts}       
\usepackage{amssymb}        
\usepackage{nicefrac}       
\usepackage{microtype}      
\usepackage[dvipsnames]{xcolor} 
\usepackage{multirow}
\usepackage{makecell}
\usepackage[capitalize,noabbrev]{cleveref}
\usepackage{subcaption}
\usepackage{array}
\usepackage{tabularx, booktabs, array, colortbl, xcolor}

\usepackage{tikz}
\usetikzlibrary{positioning, shapes, arrows.meta, fit, backgrounds, calc}

\definecolor{layerA}{RGB}{214,234,248}
\definecolor{layerB}{RGB}{212,239,223}
\definecolor{layerC}{RGB}{253,235,208}
\definecolor{layerD}{RGB}{240,219,255}
\definecolor{caipblue}{RGB}{31,97,141}
\definecolor{caipgreen}{RGB}{30,132,73}
\definecolor{caippurple}{RGB}{100,60,160}
\definecolor{caipred}{RGB}{180,60,30}
\definecolor{ifacecolor}{RGB}{180,70,40}

\begin{document}

\title{A Standards-Aligned Coordination Framework for Edge-Enhanced Collaborative Healthcare in 6G Networks}

\author{Liuwang Kang, Fan Wang, Yuzhang Huang, Shang Yan, Jianbin Zheng, Wenbin Lei, Konstantin Yakovlev, Jie Tang, Shaoshan Liu

\thanks{Liuwang Kang, Fan Wang (co-corresponding author), and Shaoshan Liu (co-corresponding author) are with the Shenzhen Institute of Artificial Intelligence and Robotics for Society (AIRS), Shenzhen, China. 
Yuzhang Huang is with the Department of Otolaryngology–Head and Neck Surgery, United Family Healthcare, Beijing, China, and also affiliated with the Chinese University of Hong Kong Medical School, Hong Kong, China.
Shang Yan is with Shenzhen Children’s Hospital, Shenzhen, China.
Jianbin Zheng is with the National Children’s Medical Center, Guangzhou Medical University Affiliated Women and Children’s Medical Center, Guangzhou, China.
Wenbin Lei is with the Otorhinolaryngology Hospital, the First Affiliated Hospital of Sun Yat-Sen University, Guangzhou, China.
Konstantin Yakovlev is with the National Research University Higher School of Economics and the Federal Research Center “Computer Science and Control” of Russian Academy of Sciences, Moscow, Russia.
Jie Tang is with the School of Computer Science and Engineering, South China University of Technology, Guangzhou, China.}
}

\IEEEpubid{\parbox{\columnwidth}{\vspace*{16pt}\raggedright\footnotesize
Preprint. This work has been submitted to the IEEE for possible publication.
Copyright may be transferred without notice, after which this version may no longer be accessible.}\hspace{\columnsep}\makebox[\columnwidth]{}}

\maketitle

\begin{abstract}

Mission-critical healthcare applications including real-time intensive care monitoring, ambulance-to-hospital orchestration, and distributed medical imaging inference require workflow-level, time-bounded coordination across heterogeneous devices, edge servers, and network control entities. While current 3GPP and O-RAN standards excel at per-device control and quality-of-service enforcement, they do not natively expose abstractions for workflow-level coordination under strict clinical timing constraints, leaving this capability to fragile, application-specific overlays. This article outlines the Collective Adaptive Intelligence Plane (CAIP) as a standards-aligned coordination framework that addresses this abstraction gap without introducing new protocol layers. CAIP is realized through minimal, backward-compatible coordination profiles anchored to existing RRC, QoS/SDAP, and O-RAN E2 interfaces, enabling workflow-scoped coordination context binding, deadline-aware coordination pacing, semantic flow association, and privacy-preserving data locality across distributed clinical entities. We analyze the structural limitations of existing standards, present a concrete interface mapping to 3GPP and O-RAN mechanisms, illustrate deployment through a representative ICU coordination scenario, and outline a phased standardization roadmap from proof-of-concept xApp deployment to AI-native 6G specification evolution. The proposed framework is incrementally deployable on current 5G Advanced infrastructure and provides a principled migration path toward workflow-level coordination abstraction as a first-class capability in future 6G healthcare networks.

\end{abstract}


\section{Introduction}

The evolution toward AI-native 6G is reshaping wireless systems into distributed, edge-enhanced computing environments. Intelligence is increasingly embedded across radio access networks, edge platforms, and cloud services to support adaptive resource allocation and real-time inference~\cite{xu2021edge}. While these advances significantly improve per-UE performance and flow-level quality-of-service (QoS) enforcement, current standardization efforts remain primarily centered on per-flow optimization mechanisms within established control hierarchies.

Healthcare networks expose a structural mismatch between such flow-centric abstractions and real service requirements. Clinical service delivery is inherently \emph{workflow-structured}, involving stateful decision graphs that span \emph{patient-side sensing}, \emph{edge analysis}, \emph{clinical sites}, \emph{mobile care nodes}, and \emph{expertise support} across administrative and regulatory domains bounded by a \emph{health data and governance boundary}. These interactions are \emph{multi-actor}, \emph{deadline-driven}, and \emph{privacy-constrained}, where the relevant performance objective is workflow completion within bounded clinical time budgets rather than individual link latency. 
For example, in emergency response coordination, an anomaly detected by a wearable sensor may trigger edge validation, ambulance dispatch, hospital preparation, and specialist consultation across multiple administrative domains. Each stage depends on prior workflow state confirmation and must be completed within clinically defined deadlines while respecting governance constraints on data exchange. Such processes cannot be reduced to independent traffic flows without losing their structural semantics, as the workflow objective spans multiple heterogeneous actors and control domains.

Existing 3GPP and O-RAN architectures provide robust mechanisms for connectivity, mobility management, and QoS guarantees. However, they do not define standardized abstractions for expressing collective workflow objectives, binding task-scoped coordination contexts, or enforcing bounded coordination cycles across heterogeneous control tiers. Their control abstractions are fundamentally flow-centric, focusing on session management and resource allocation rather than workflow-state representation and cross-actor objective alignment. As a result, healthcare-grade collaboration must rely on application-layer orchestration outside the standardized control plane, limiting interoperability and weakening deterministic performance guarantees. In the absence of coordination semantics at the control-plane level, workflow-driven healthcare objectives cannot be consistently represented within standardized network procedures.

To address this structural gap, this article outlines the \emph{Collective Adaptive Intelligence Plane (CAIP)} as a standards-aligned coordination framework for AI-native 6G healthcare networks~\cite{wang2025empowering}. Rather than introducing a new protocol stack, CAIP provides a coordination abstraction that aligns healthcare workflow semantics with existing control hierarchies through bounded interaction cycles, collective objective indicators, and metadata-driven context propagation. It enables deadline-aware coordination and governance-constrained collaboration while remaining compatible with current 3GPP and O-RAN interfaces. In this sense, CAIP is not merely an efficiency enhancement, but a structural extension required to support workflow-level edge-enhanced collaborative computing across patient-side sensing, edge analysis, clinical sites, mobile care nodes, and expertise support domains.

The remainder of this article first analyzes the structural limitations of existing standards in supporting workflow-driven healthcare coordination. It then presents CAIP as a standards-aligned coordination layer and details its interface mapping to 3GPP and O-RAN architectures. Finally, representative healthcare deployment scenarios and standardization pathways are discussed.

\begin{figure*}[t]
\centering
\includegraphics[width=\textwidth]{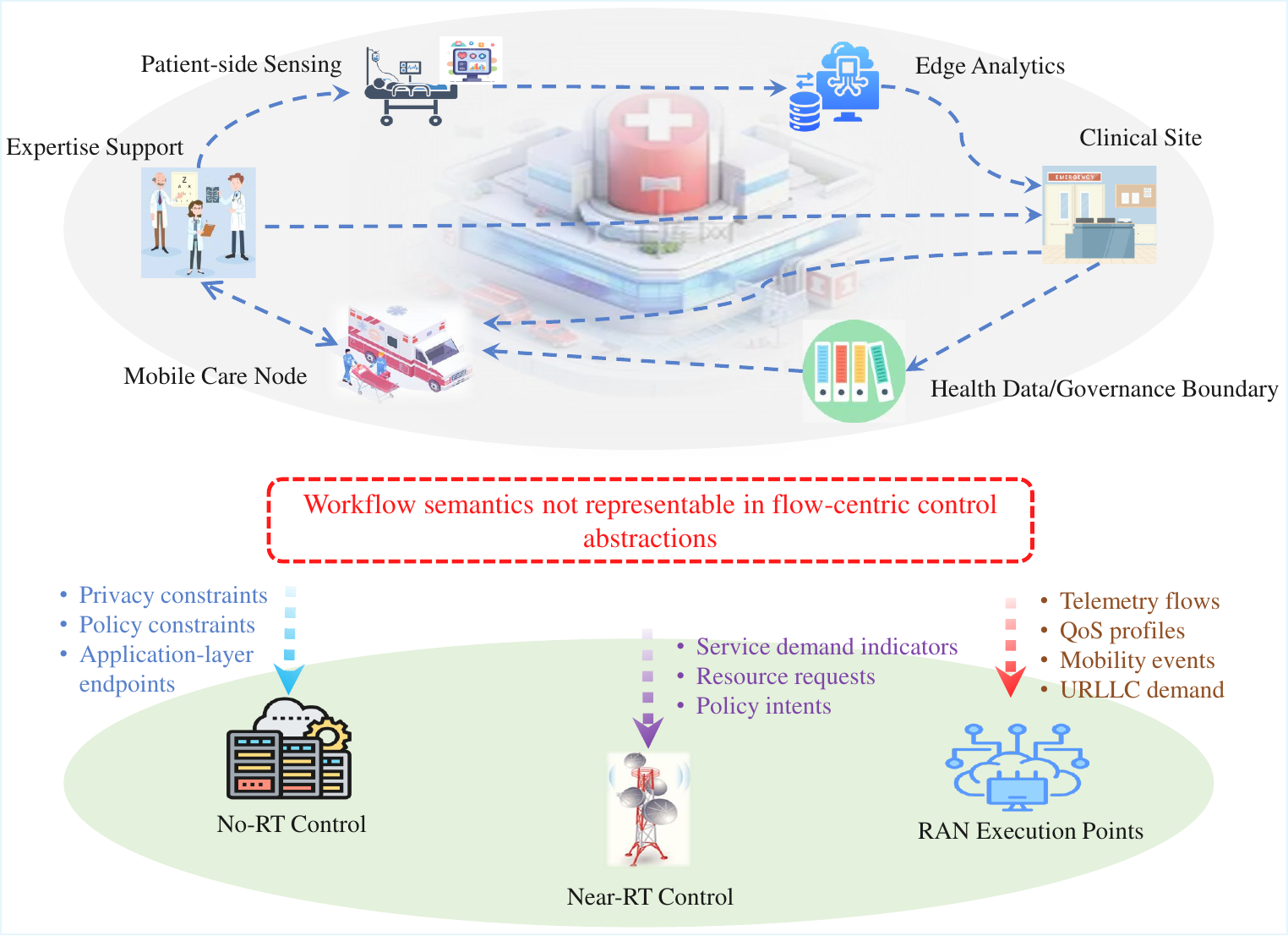}
\caption{Structural mismatch between workflow-level healthcare service delivery and flow-centric control abstractions in existing 3GPP and O-RAN architectures. The absence of workflow-level coordination semantics indicates the need for a coordination abstraction layer.}
\label{fig:workflow_mismatch}
\end{figure*}

\section{Limitations of Existing Standards for Healthcare-Grade Coordination}
\label{sec:limitations}

Current 3GPP and O-RAN specifications provide a robust baseline for connectivity, radio optimization, and QoS enforcement~\cite{3gpp2023nr,tripathi2025fundamentals}. RRC procedures support session and mobility management, MAC scheduling and URLLC mechanisms enable tight link-layer control~\cite{3gpp2023nr}, SDAP supports QoS flow mapping~\cite{3gpp-ts33501}, and near-RT RIC control loops allow timely policy steering. These capabilities are essential for latency-sensitive healthcare services. However, they predominantly treat performance as a per-UE or per-cell optimization problem. In healthcare-grade collaboration, the primary objective is often not individual link performance, but workflow completion under clinical deadlines, where multiple structural roles must align actions and resource usage within bounded coordination cycles.

A fundamental limitation therefore lies in the absence of a native workflow-level coordination abstraction. Healthcare workflows routinely span heterogeneous structural roles, including patient-side sensing, edge analysis, clinical sites, mobile care nodes, and expertise support, often operating across regulated health data and governance boundaries~\cite{liu2022rise, liu2022autonomous}. Although existing standards can establish and prioritize individual bearers~\cite{3gpp-ts23501}, they do not provide a standardized construct to represent a shared coordination context (e.g., a task group identifier, a common deadline budget, or an explicit workflow state) that remains consistently visible across control tiers. As a consequence, multi-entity coordination is typically delegated to application-layer orchestrators operating outside standardized RAN control semantics. This separation weakens the coupling between collective decisions and radio or transport enforcement, and may result in latency accumulation across independently managed sessions, unstable prioritization under competing clinical tasks, and difficulty in enforcing coordination decisions within predefined control-cycle budgets under fluctuating load conditions.

A related limitation is the dominance of node-centric objectives within standardized control loops. gNB scheduling, radio resource management (RRM), and near-RT RIC applications are designed to optimize local indicators such as throughput, packet delay, and link reliability within well-defined time-scale constraints~\cite{3gpp2023nr}. Healthcare-grade coordination, however, may require temporary system-level tradeoffs to satisfy a collective deadline. For example, synchronizing activity across patient-side sensing and edge analysis, reserving resources for time-critical validation flows at clinical sites, or coordinating task transitions involving mobile care nodes and expertise support. Current specifications offer limited means to express and enforce such collective objectives within the standardized control hierarchy. As a result, network behavior remains fragmented across independently optimized entities, even when clinical outcomes depend on coordinated action.

A further limitation concerns the lack of semantic binding between control decisions and service workflows. While the 3GPP QoS framework enables traffic differentiation through QoS flows and 5QI mappings~\cite{lei20215g}, these mechanisms primarily classify traffic types rather than encode workflow semantics. A clinical process may span multiple sessions and governance domains while progressing through explicit task states, such as detection, validation, and escalation. Yet the control plane does not provide a standardized method to associate resource adjustments or policy updates with a unified workflow context. Without such binding, control functions cannot systematically reason about task progression, pace coordination rounds according to deadline proximity, or adapt behavior based on workflow completion status.

Finally, healthcare deployments are shaped by privacy-driven cross-domain constraints that influence how coordination can be realized. Although standards define authentication, encryption, and secure transport primitives~\cite{3gpp-ts33501}, they do not structure coordination semantics for regulated, multi-administrative collaboration environments in which raw clinical data must remain within a health data and governance boundary and only minimal coordination metadata may traverse domains. Consequently, interoperable coordination logic is often implemented in isolated platforms using proprietary conventions. This reduces portability across operators and limits the integration of privacy-aware coordination behavior into standardized control loops.

In summary, existing 3GPP and O-RAN standards provide strong foundations for connectivity, mobility management, and localized optimization. However, they do not natively support bounded-time multi-actor coordination, collective objective expression across heterogeneous structural roles, or workflow-level semantic binding between service intent and control actions. These limitations become particularly critical in healthcare environments, where collaboration spans institutional and governance boundaries and must operate under enforceable latency and privacy constraints. As illustrated in Fig.~\ref{fig:workflow_mismatch}, healthcare service delivery forms a stateful, multi-actor workflow governed by bounded clinical deadlines, whereas current control hierarchies remain primarily flow-centric. This abstraction mismatch prevents workflow-level objectives from being represented within standardized control procedures. Addressing this gap requires a coordination abstraction capable of aligning workflow semantics with existing control mechanisms without altering underlying protocol structures.

\begin{figure*}[t]
\centering
\includegraphics[width=\textwidth]{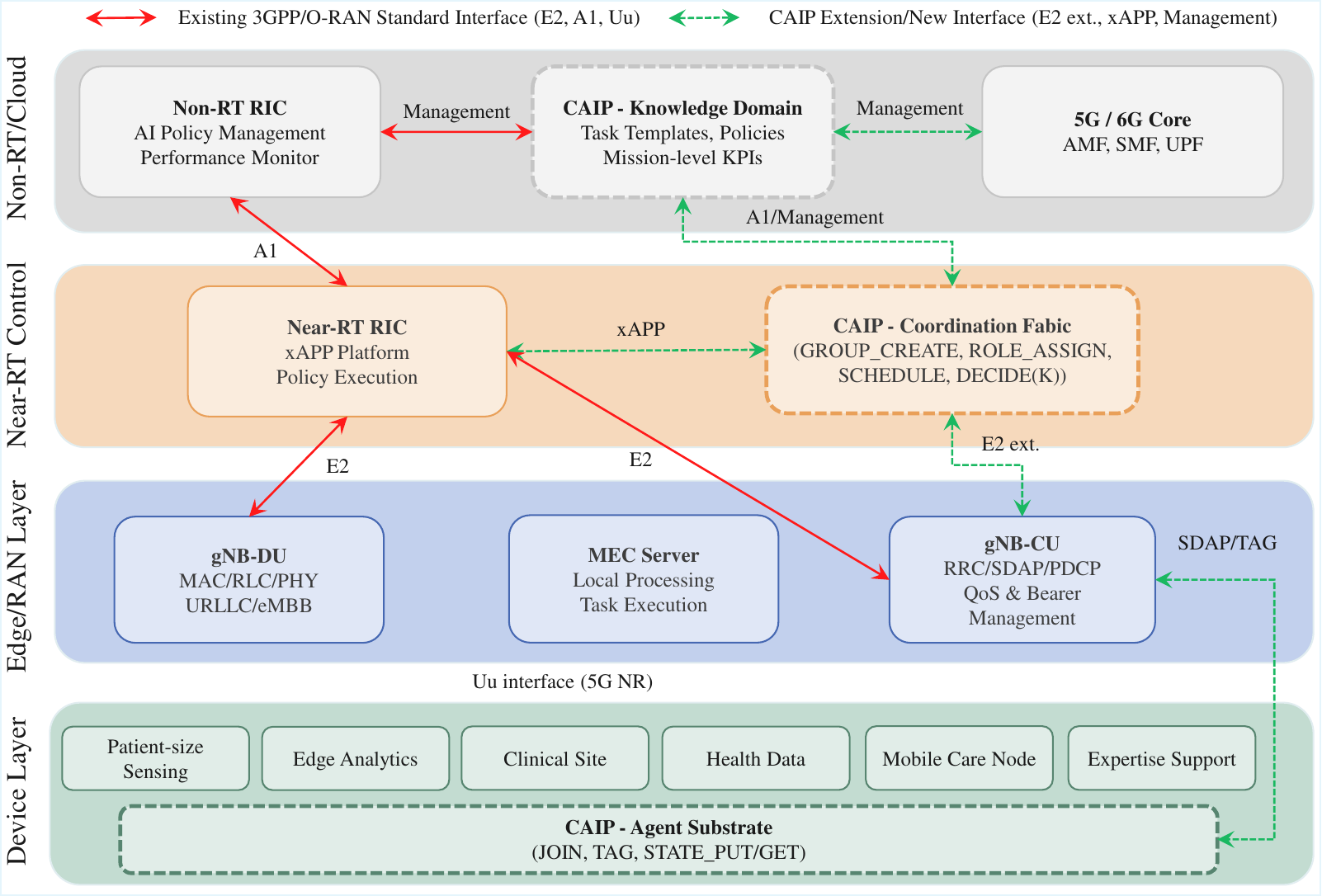}
\caption{Standards-aligned architecture of CAIP for 6G healthcare coordination. The Agent Substrate (device layer), Coordination Fabric (near-RT RIC), and Knowledge Domain (non-RT cloud) align with existing 3GPP and O-RAN control tiers through Uu, E2, A1, and management interfaces. Solid arrows denote standardized interfaces; dashed blue arrows indicate optional coordination extensions.}
\label{fig:caip_arch}
\end{figure*}

\section{CAIP as a Standards-Aligned Coordination Framework}
\label{sec:caip}

The coordination gaps identified in Section~\ref{sec:limitations} arise from the absence of a standardized construct for bounded-time, workflow-level coordination within the existing 3GPP and O-RAN control hierarchy. To address this structural limitation, we outline the \emph{CAIP} as a standards-aligned coordination framework that augments current architectures while preserving established radio procedures. Rather than introducing new protocol layers, CAIP organizes workflow-level coordination semantics across existing control tiers and time scales using minimal extensions to standardized interfaces.

As illustrated in Fig.~\ref{fig:caip_arch}, CAIP aligns with three control tiers already present in existing communication systems. At the device and edge level, the \emph{Agent Substrate} supports capability advertisement, lightweight workflow-state signaling~\cite{fan2025putting}, and association of service flows with a coordination context through existing QoS and SDAP mapping procedures. This substrate enables structural roles such as patient-side sensing and edge analysis to participate in coordinated workflows without exposing sensitive healthcare data beyond the originating \emph{health data and governance boundary}. Only coordination metadata is exchanged at this tier.

Within the near-RT domain, the \emph{Coordination Fabric} operates as a RIC application that structures task grouping, role alignment across heterogeneous structural domains, and bounded coordination cycles. It interfaces with network elements through E2 without altering MAC or PHY operations. This tier enables collective workflow objectives to be evaluated and enforced within standardized near-RT control envelopes~\cite{oran-arch}, thereby aligning workflow semantics with existing scheduling and resource management procedures.

At the non-RT tier, the \emph{Knowledge Domain} maintains workflow templates, coordination policies, and long-horizon artifacts provisioned through A1 and management interfaces. This layer ensures alignment between healthcare workflow objectives and network policy configurations across administrative and governance boundaries~\cite{letaief2021edge,letaief2019roadmap}, supporting multi-domain deployments while preserving interoperability.

CAIP coordination executes within predefined control cycles aligned with standardized latency budgets~\cite{oran-arch}. Instead of relying on centralized global optimization, collective behavior is structured through finite interaction rounds embedded in near-RT control loops. Coordination context identifiers, collective objective indicators, and deadline attributes are conveyed using optional signaling attributes anchored to existing interfaces, ensuring compatibility with scheduling and resource management mechanisms.

\begin{table}[!t]
\renewcommand{\arraystretch}{1.0}
\caption{Mapping of CAIP Components to Standardized Control Tiers and Interfaces.}
\label{tab:caip_mapping}
\centering
\small
\begin{tabularx}{\columnwidth}{@{}lXl@{}}  
\toprule
\textbf{Component} & \textbf{Host Tier} & \textbf{Primary Interface} \\
\midrule
Agent Substrate      & Device/gNB-DU      & RRC, SDAP (Uu) \\
Coordination Fabric  & Near-RT RIC          & E2 (E2SM ext.) \\
Knowledge Domain     & Non-RT RIC/Cloud   & A1, Management \\
\bottomrule
\end{tabularx}
\end{table}

Table~\ref{tab:caip_mapping} summarizes the mapping between CAIP components, their host control tiers, and the primary standardized interfaces through which coordination semantics are realized. Each component is implemented through optional extensions to existing control semantics, enabling incremental deployment. Networks that do not implement CAIP continue to operate unchanged, while CAIP-enabled deployments gain structured, latency-bounded workflow-level coordination capabilities within the standardized control hierarchy. CAIP does not replace application-layer orchestration platforms; rather, it provides a coordination abstraction within the standardized control plane, allowing healthcare workflow objectives to be consistently reflected in radio and policy control loops while preserving backward compatibility.

\begin{table*}[t]
\caption{Profile-based realization of CAIP coordination semantics across existing 3GPP and O-RAN interfaces. All enhancements are optional and backward-compatible.}
\label{tab:caip_minimal_extension}
\centering
\setlength{\tabcolsep}{7pt}
\renewcommand{\arraystretch}{1.4}
\begin{tabularx}{\textwidth}{
  >{\raggedright\arraybackslash}p{2.2cm}
  >{\raggedright\arraybackslash}X
  >{\raggedright\arraybackslash}X}

\toprule
\textbf{Interface} &
\textbf{CAIP Context Realization} &
\textbf{Standards Vehicle / Impact} \\
\midrule

\textbf{RRC} &
Optional \textit{GroupID} and \textit{TimeBudget} carried in \texttt{RRCReconfiguration}. &
Optional IE defined within TS~38.331 extension mechanisms; no change to RRC procedures, with normative specification updates limited to optional attribute definitions. \\

\textbf{SDAP} &
Workflow context associated with QoS flows through QoS policy mapping and coordination-context interpretation at higher control layers. &
Profile-based realization using existing QoS flow mechanisms under TS~37.324.\\

\textbf{E2 (E2SM)} &
Coordination state and deadline indicators included in E2SM profiles. &
Optional E2SM fields; aligned with O-RAN WG3 extensibility principles. \\

\textbf{A1} &
Healthcare task descriptors provisioned as A1 policy objects. &
Existing A1 policy framework; no new procedures required. \\

\textbf{MAC / PHY} &
No modification. &
Fully compliant with existing 3GPP specifications. \\

\bottomrule
\end{tabularx}
\end{table*}

\section{Interface Mapping to 3GPP and O-RAN Architectures}

Table~\ref{tab:caip_minimal_extension} summarizes how CAIP coordination semantics can be realized through minimal, backward-compatible extensions to existing 3GPP and O-RAN interfaces. Rather than introducing new protocol layers or redefining control procedures, CAIP associates workflow-level coordination context with already standardized signaling and policy mechanisms. All enhancements are optional and profile-based, and lower-layer radio procedures remain unchanged.

At the UE control layer, coordination context can be conveyed using optional information elements carried in existing \texttt{RRCReconfiguration} messages, consistent with 3GPP TS~38.331~\cite{3gpp2020nr}. A \textit{GroupID} parameter associates a UE with a healthcare workflow instance spanning structural roles such as patient-side sensing and edge analysis, while a \textit{TimeBudget} attribute indicates the bounded participation window for that task. Since RRC specifications already allow optional information elements and extension containers, legacy devices that do not recognize these parameters ignore them without affecting session continuity. No modification to the RRC state machine or message flow is required at runtime; however, the introduction of additional optional information elements would necessitate corresponding specification updates within the existing extension framework.

Workflow-to-QoS association can be achieved at the SDAP layer without redefining bearer structures. CAIP does not introduce new 5QI identifiers or data radio bearer types. Instead, coordination context may be realized through QoS flow association and policy interpretation at higher layers, allowing workflow participation across patient-side sensing, clinical sites, and mobile care nodes to be consistently mapped to existing QoS mechanisms without modifying SDAP header structure. This enables multiple traffic streams participating in the same healthcare workflow to be consistently prioritized under existing QoS rules. Because SDAP already maps QoS flows to radio bearers, the CAIP enhancement operates strictly at the coordination-context level and does not alter bearer establishment or scheduling logic.

Near-RT coordination is realized within O-RAN near-RT RIC applications through the E2 interface. Coordination state indicators, collective objective attributes, and deadline-related parameters can be encoded as optional fields within E2 Service Model (E2SM) profiles, in alignment with O-RAN WG3 extensibility principles. Subscription procedures, reporting intervals, and control message timing remain unchanged. By confining workflow-level coordination semantics to E2SM-level extensions, CAIP enables bounded coordination cycles within the established near-RT control envelope without introducing new control-plane protocols.

At the non-RT tier, collective objectives and healthcare workflow templates are provisioned through the A1 interface. CAIP-compatible policy objects may include task descriptors, deadline thresholds, escalation conditions, and governance-aware coordination constraints using existing A1 policy frameworks. The non-RT RIC continues to operate on longer optimization cycles, distributing policy parameters that guide near-RT coordination decisions across structural roles and administrative domains. This preserves the standardized separation between long-horizon analytics and time-sensitive execution while respecting health data and governance boundaries.

CAIP does not modify MAC or PHY procedures. URLLC scheduling, HARQ timing, link adaptation, and retransmission mechanisms remain fully compliant with current 3GPP specifications. Coordination semantics are introduced strictly above the radio resource management layer, preserving the integrity of the standardized control hierarchy. Because all extensions are optional and profile-based, incremental deployment is feasible: operators may introduce CAIP-enabled RIC applications and network-side coordination logic without requiring simultaneous UE upgrades, provided that workflow-level coordination semantics are interpreted within network control entities. Vendors may implement E2SM or A1 profile extensions while remaining compliant with baseline specifications.

Through this interface-anchored realization, CAIP embeds healthcare workflow semantics into existing control mechanisms while preserving interoperability and backward compatibility. Initial deployment may operate entirely within network-side RIC and control-plane entities without requiring UE upgrades, where coordination semantics are enforced through scheduling and policy decisions. End-to-end UE-aware coordination, however, requires support for the corresponding optional signaling attributes and can be introduced in later specification phases.

\begin{figure*}[t]
\centering
\includegraphics[width=\textwidth]{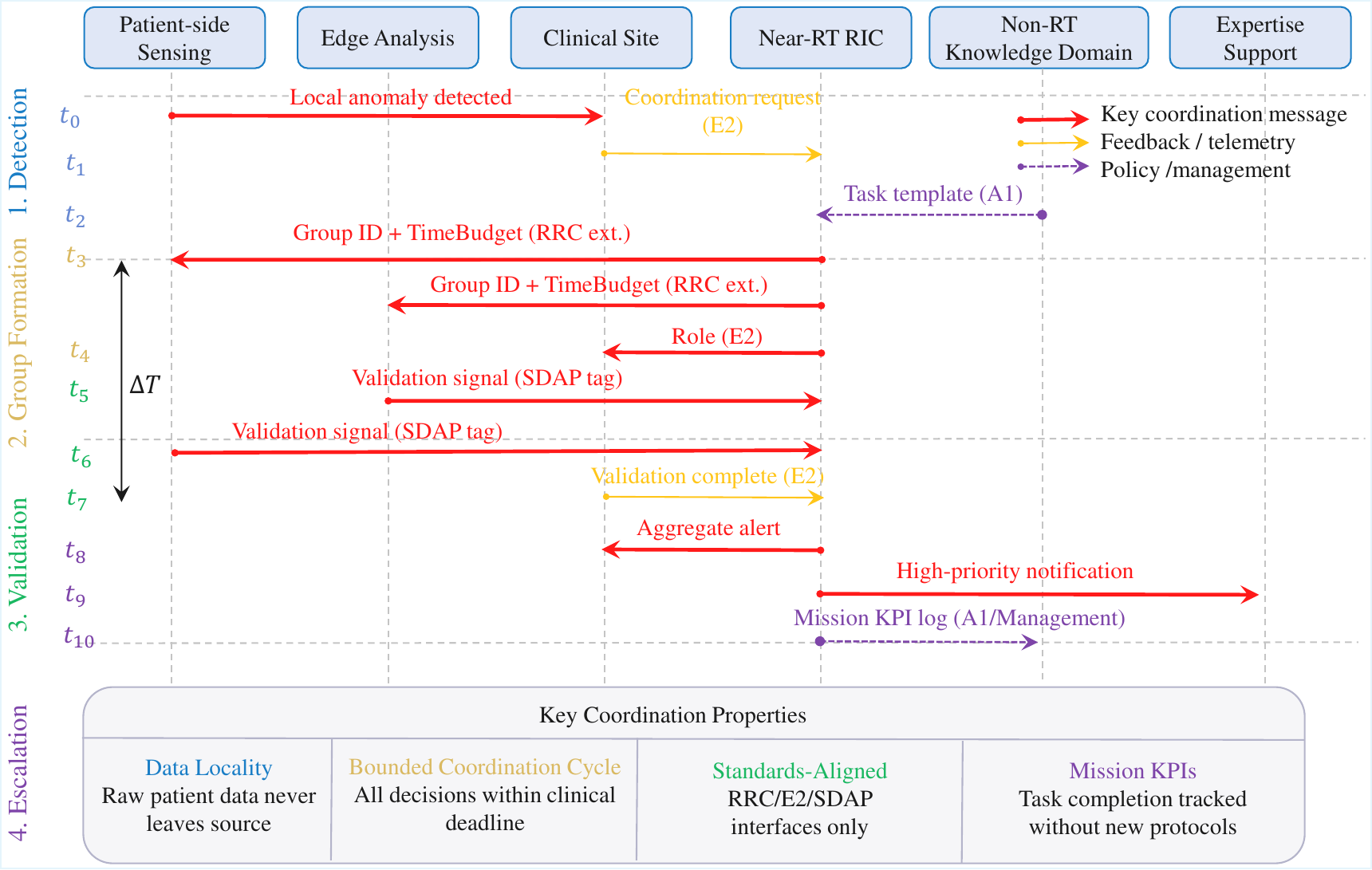}
\caption{Representative CAIP-enabled ICU anomaly validation workflow in a 6G healthcare network. 
Six structural roles coordinate across four stages: anomaly detection, task-oriented group formation via RRC and E2 signaling, bounded-time validation using SDAP-associated flows, and role-based escalation with KPI recording. Only coordination metadata is exchanged through standardized control interfaces; raw patient data remains local within each health data and governance boundary.}
\label{fig:healthcare_workflow}
\end{figure*}

\section{Privacy-Preserving Edge Collaboration in Healthcare}

Healthcare coordination operates under strict regulatory and organizational constraints. Patient information is governed by data-protection regulations, institutional trust boundaries, and cross-domain governance policies defined by distinct \emph{health data and governance boundaries}. Although 3GPP and O-RAN specifications define secure transport, authentication, and encryption mechanisms, they do not explicitly structure how workflow-level, multi-actor coordination can be realized across administrative domains while preserving data locality. In healthcare-grade deployments, the challenge extends beyond secure communication to enabling predictable, bounded-time collaboration without centralizing protected data.

Figure~\ref{fig:healthcare_workflow} illustrates how CAIP supports privacy-aware coordination across distributed structural roles, including patient-side sensing, edge analysis, clinical sites, mobile care nodes, and expertise support domains. Raw patient data remains confined to its originating health data and governance boundary—for example, within an ICU edge analysis system or a clinical imaging platform—while only coordination metadata traverses standardized control interfaces. These metadata elements include workflow identifiers, deadline indicators, role-alignment signals, completion status flags, and escalation indicators. Diagnostic images, biometric streams, and clinical records are not propagated through coordination channels. At the non-RT tier, workflow templates and task descriptors may be provisioned through A1 interfaces within the Non-RT Knowledge Domain, consistent with the standardized separation between long-horizon policy provisioning and time-sensitive coordination.

Coordination proceeds through bounded coordination cycles that exchange abstracted workflow-state indicators rather than application-layer content. In the anomaly validation workflow shown in Fig.~\ref{fig:healthcare_workflow}, participating structural roles signal role alignment via E2-mediated coordination context exchange and indicate deadline proximity using standardized control-plane mechanisms. These signals are sufficient to enable near-RT resource alignment within RRC-, SDAP-, and E2-mediated control loops. By operating on workflow-level state summaries instead of aggregated datasets, CAIP preserves data locality and remains consistent with regulated healthcare deployment requirements.

CAIP does not introduce new security procedures. Coordination identifiers and collective objective indicators are transported through the same authenticated and encrypted control-plane channels defined in existing 3GPP and O-RAN specifications. Because CAIP relies on optional signaling attributes and policy objects anchored to standardized interfaces, it inherits established trust models and credential management mechanisms without requiring new protocol stacks.

This metadata-driven coordination model also supports federated, cross-domain collaboration. Each administrative domain retains authority over local resource enforcement within its governance boundary, while bounded coordination cycles align decision timing across heterogeneous structural roles. Rather than centralizing control logic, CAIP synchronizes workflow-level coordination within the standardized hierarchy, enabling interoperable healthcare workflows that respect privacy constraints and institutional boundaries. Mission-level KPI indicators may be recorded through A1 or management interfaces to track workflow completion status without exposing protected clinical data.

\begin{figure*}[htp]
\centering
\includegraphics[width=\textwidth]{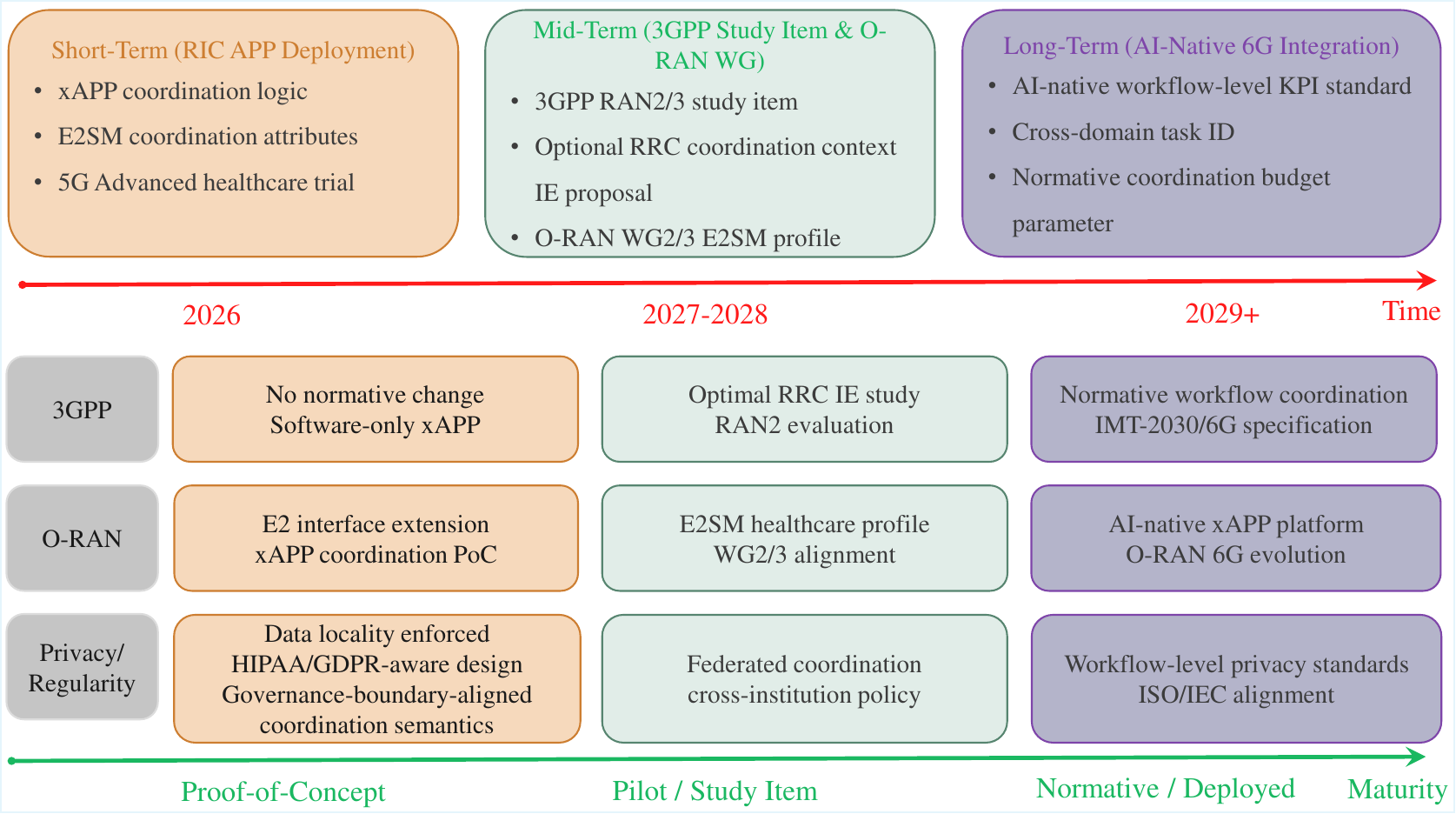}
\caption{Phased roadmap for CAIP standardization and deployment aligned with 3GPP and O-RAN evolution. The timeline outlines incremental realization from software-level coordination abstraction to optional specification support and long-term vertical workflow profiles. Backward compatibility is preserved throughout; normative enhancements are limited to later phases.}
\label{fig:roadmap}
\end{figure*}

\section{Standardization and Deployment Roadmap}

The integration of CAIP into AI-native 6G systems can follow a non-disruptive and phased evolution path aligned with established 3GPP and O-RAN standardization processes. Rather than requiring architectural redesign, workflow-level coordination semantics can be progressively introduced within existing control hierarchies. Fig.~\ref{fig:roadmap} summarizes this roadmap and its alignment with expected industry milestones.

In the initial phase, CAIP capabilities can be realized within existing near-RT RIC platforms as optional coordination functions implemented through xApps and policy modules. At this stage, workflow-level coordination context identifiers and bounded coordination cycle semantics may be encoded using vendor-specific profiles within permissible interface fields, without modifying standardized control procedures. 

This approach enables proof-of-concept deployment, interoperability experimentation, and vertical-industry validation. In healthcare deployments, such validation should ultimately connect coordination KPIs (e.g., bounded coordination cycles and escalation latency) to workflow-level service outcomes within governance-constrained environments. For example, anomaly validation workflows spanning patient-side sensing, edge analysis, clinical sites, and expertise support domains may be evaluated against bounded clinical deadlines and coordination consistency across health data and governance boundaries~\cite{huang2024health}.

Building on operational experience, a second phase may formalize coordination semantics as optional information elements or policy attributes within relevant specifications. For example, standardized definitions of coordination state indicators, workflow-level objective descriptors, or deadline attributes may be incorporated into E2 service models or A1 policy formats through study-item and work-item processes. These additions would remain backward compatible and optional, allowing legacy devices and networks to continue operating unchanged while providing structured support for coordination-aware deployments.

In a longer-term phase, regulated verticals such as healthcare may adopt standardized workflow coordination profiles. Such profiles would define permissible timing constraints, workflow context parameters, governance-boundary interpretation rules, and interoperability requirements for multi-operator environments. Rather than redefining radio procedures, these profiles would clarify how coordination context identifiers and objective indicators are interpreted across administrative and governance domains, supporting federated collaboration while preserving operator autonomy.

Throughout all phases, interoperability and vendor neutrality are preserved by anchoring coordination semantics to existing interfaces rather than introducing new transport protocols or bearer types. Conformance testing can leverage established multi-vendor RIC platforms and standardized verification procedures. As 6G architectures continue to evolve toward AI-native operation, structured workflow-level coordination semantics can transition from experimental capabilities to formally recognized control-plane constructs, enabling healthcare-grade collaboration while maintaining architectural stability.

\section{Conclusion}

Healthcare-grade 6G services require a workflow-level, bounded-time coordination abstraction across distributed devices, edge systems, and control-plane entities—capabilities that are not natively structured within current 3GPP and O-RAN specifications. This article has outlined CAIP as a standards-aligned coordination framework that embeds workflow-level coordination semantics into existing control hierarchies through optional, backward-compatible interface extensions, operating strictly above the radio resource layer while preserving established time-scale separation. By anchoring coordination semantics to standardized interfaces, CAIP supports governance-boundary-aligned, privacy-aware multi-entity collaboration without architectural disruption. The phased evolution path described herein demonstrates that workflow-level coordination capabilities can evolve incrementally—from software-level realization to optional specification support and ultimately vertical coordination profiles—establishing a structured foundation for interoperable AI-native 6G healthcare networks.

\bibliographystyle{unsrtnat}
\bibliography{main}

@article{liu2022rise,
  title={Rise of the automotive health-domain controllers: Empowering healthcare services in intelligent vehicles},
  author={Liu, Shaoshan and Huang, Yuzhang and Kong, Ao and Tang, Jie and Liu, Xue},
  journal={IEEE Internet of Things Journal},
  volume={9},
  number={24},
  pages={24882--24889},
  year={2022},
  publisher={IEEE}
}

@article{wang2025empowering,
  title={Empowering virtual agents with intelligent systems},
  author={Wang, Fan and Liu, Shaoshan},
  journal={Communications of the ACM},
  volume={68},
  number={8},
  pages={8--9},
  year={2025},
  publisher={ACM New York, NY, USA}
}

@article{liu2022autonomous,
  title={Autonomous mobile clinics},
  author={Liu, Shaoshan and Kong, Ao and Huang, Yuzhang and Liu, Xue},
  journal={Bulletin of the World Health Organization},
  volume={100},
  number={9},
  pages={527},
  year={2022}
}

@article{huang2024health,
  title={Health satisfaction outcome from integrated autonomous mobile clinics},
  author={Huang, Yuzhang and Liu, Shaoshan and Pan, Zhongying and Wu, Carl and Chiu, Herng-Chia and Liu, Xue and Shi, Leiyu},
  journal={Scientific reports},
  volume={14},
  number={1},
  pages={24878},
  year={2024},
  publisher={Nature Publishing Group UK London}
}

@misc{oran-arch,
  author    = {{O-RAN Alliance}},
  title     = {O-RAN Architecture Description and White Papers},
  year      = {2018--2024},
  note      = {Describes Near-RT RIC (E2), Non-RT RIC/SMO (A1/O1), O-Cloud, and roles}
}

@article{letaief2021edge,
  title={Edge artificial intelligence for 6G: Vision, enabling technologies, and applications},
  author={Letaief, Khaled B and Shi, Yuanming and Lu, Jianmin and Lu, Jianhua},
  journal={IEEE journal on selected areas in communications},
  volume={40},
  number={1},
  pages={5--36},
  year={2021},
  publisher={IEEE}
}

@article{letaief2019roadmap,
  title={The roadmap to 6G: AI empowered wireless networks},
  author={Letaief, Khaled B and Chen, Wei and Shi, Yuanming and Zhang, Jun and Zhang, Ying-Jun Angela},
  journal={IEEE communications magazine},
  volume={57},
  number={8},
  pages={84--90},
  year={2019},
  publisher={IEEE}
}

@article{fan2025putting,
  title={Putting the Smarts into Robot Bodies},
  author={Fan, Wang and Liu, Shaoshan},
  journal={Communications of the ACM},
  volume={68},
  number={3},
  pages={6--8},
  year={2025},
  publisher={ACM New York, NY, USA}
}

@techreport{3gpp-ts23501,
  author       = {{3GPP}},
  title        = {System architecture for the 5G System (5GS)},
  institution  = {3rd Generation Partnership Project (3GPP)},
  number       = {TS 23.501 (ETSI TS 123 501)},
  version      = {V18.5.0},
  year         = {2024},
  month        = may,
  note         = {Release 18},
  url          = {https://www.etsi.org/deliver/etsi_ts/123500_123599/123501/18.05.00_60/ts_123501v180500p.pdf}
}

@misc{3gpp2023nr,
  title={NR; NR and NG-RAN Overall description; Stage-2},
  author={3GPP TS 38.300, JPSH},
  year={2023}
}

@incollection{lei20215g,
  title={5G system architecture},
  author={Lei, Wan and Soong, Anthony CK and Jianghua, Liu and Yong, Wu and Classon, Brian and Xiao, Weimin and Mazzarese, David and Yang, Zhao and Saboorian, Tony},
  booktitle={5G System Design: An End to End Perspective},
  pages={297--339},
  year={2021},
  publisher={Springer}
}

@techreport{3gpp-ts33501,
  author       = {{3GPP}},
  title        = {Security architecture and procedures for 5G system},
  institution  = {3rd Generation Partnership Project (3GPP)},
  number       = {TS 33.501},
  version      = {V18.4.0},
  year         = {2024},
  month        = mar,
  note         = {Release 18}
}

@book{tripathi2025fundamentals,
  title={Fundamentals of O-RAN},
  author={Tripathi, Nishith D and Shah, Vijay K},
  year={2025},
  publisher={John Wiley \& Sons}
}

@misc{3gpp2020nr,
  title={NR; Radio Resource Control (RRC); Protocol specification},
  author={3GPP TS 38.331, H},
  year={2020}
}

@article{xu2021edge,
  title={Edge intelligence: Empowering intelligence to the edge of network},
  author={Xu, Dianlei and Li, Tong and Li, Yong and Su, Xiang and Tarkoma, Sasu and Jiang, Tao and Crowcroft, Jon and Hui, Pan},
  journal={Proceedings of the IEEE},
  volume={109},
  number={11},
  pages={1778--1837},
  year={2021},
  publisher={IEEE}
}

\begin{IEEEbiographynophoto}{LIUWANG KANG}
received the Ph.D. degree from the University of Virginia, Charlottesville, VA, USA, in 2021. He is currently a Researcher with the Shenzhen Institute of Artificial Intelligence and Robotics for Society (AIRS), Shenzhen, China. His research interests include cyber--physical systems and 6G embodied AI. (kangliuwang@cuhk.edu.cn)
\end{IEEEbiographynophoto}

\begin{IEEEbiographynophoto}{FAN WANG}
is a research scientist at Shenzhen Institute of Artificial Intelligence and Robotics for Society, and formerly serves as a Distinguished Architect at Baidu Inc. He holds a Doctor degree of Electronic and Information Engineering from the University of Science and Technology of China (USTC), a Master degree from the engineering school at the University of Colorado at Boulder, and a Bachelor degree from USTC. His research area include large-scale pretraining, generative models, LLM and autonomous driving. He is currently interested in large-scale meta-learning, in-context learning and self-evolving agents. (fanwang.px@gmail.com)
\end{IEEEbiographynophoto}

\begin{IEEEbiographynophoto}{YUZHANG HUANG,}
M.D. (Member, IEEE) is currently the Chair and Professor of the Department of Otolaryngology–Head and Neck Surgery, United Family Healthcare. He is also an Affiliated Professor with the Chinese University of Hong Kong Medical School. (huang.yuzhang@ufh.com.cn)
\end{IEEEbiographynophoto}

\begin{IEEEbiographynophoto}{SHANG YAN,}
M.D. is currently an Associate Chief Physician and a Master’s Supervisor with Shenzhen Children’s Hospital, China. His research interests include hereditary hearing loss and upper airway–related diseases. (yshang0616@163.com)
\end{IEEEbiographynophoto}

\begin{IEEEbiographynophoto}{JIANBIN ZHENG,}
M.D. is currently a Chief Physician and Master's Supervisor at the National Children's Medical Center, Guangzhou Medical University Affiliated Women and Children's Medical Center, Guangzhou, China. Dr. Zheng is a recipient of the First Prize of Guangdong Medical Science and Technology Award. (zhengjianbin@gwcmc.org)
\end{IEEEbiographynophoto}

\begin{IEEEbiographynophoto}{WENBIN LEI,}
M.D. is Dean of Otorhinolaryngology Hospital, the First Affiliated Hospital, Sun Yat-Sen University, National Key Department of Otorhinolaryngology of People's Republic of China. (leiwb@mail.sysu.edu.cn)
\end{IEEEbiographynophoto}

\begin{IEEEbiographynophoto}{KONSTANTIN YAKOVLEV}
is an Associate Professor at National Research University Higher School of Economics and Department Head of Joint Department with the Federal Research Center 'Computer Science and Control' of the Russian Academy of Sciences. (yakovlev.ks@gmail.com)
\end{IEEEbiographynophoto}

\begin{IEEEbiographynophoto}{JIE TANG}
(Senior Member, IEEE) is currently an associate professor in School of Computer Science and Engineering of South China University of Technology, Guangzhou, China. Dr. Tang is mainly doing research on Computing Systems for Autonomous Machines and Multimodal Large Model.(cstangjie@scut.edu.cn)
\end{IEEEbiographynophoto}

\begin{IEEEbiographynophoto}{SHAOSHAN LIU}
is Director of Embodied AI at Shenzhen Institute of Artificial Intelligence and Robotics for Society (AIRS). He is an Elected Member of the Global Young Academy, an IEEE Senior Member, an IEEE Computer Society Distinguished Speaker, an ACM Distinguished Speaker, and an agenda contributor of the World Economic Forum. (shaoshanliu@cuhk.edu.cn)
\end{IEEEbiographynophoto}

\end{document}